\documentclass{pasj00}

\begin{document}
\SetRunningHead{H. Toujima et al.}{NH$_3$ and H$_2$O maser emissions in NGC 7000}
\Received{2009/10/06}
\Accepted{2011/07/11}

\title{Propagation of Highly Efficient Star Formation in NGC 7000}

\author{Hideyuki \textsc{Toujima},\altaffilmark{1} 
        Takumi \textsc{Nagayama},\altaffilmark{2}
        Toshihiro \textsc{Omodaka},\altaffilmark{1, 3}
        Toshihiro \textsc{Handa},\altaffilmark{4*} \\
        Yasuhiro \textsc{Koyama},\altaffilmark{5}
        and Hideyuki \textsc{Kobayashi}\altaffilmark{2}}
\altaffiltext{1}{Graduate School of Science and Engineering, Kagoshima University, \\
                 1-21-35 Korimoto, Kagoshima, Kagoshima 890-0065}
\altaffiltext{2}{Mizusawa VLBI Observatory, National Astronomical Observatory of Japan, \\
                 2-21-1 Osawa, Mitaka, Tokyo 181-8588}
\altaffiltext{3}{Faculty of Science, Kagoshima University, \\
                 1-21-35 Korimoto, Kagoshima, Kagoshima 890-0065}
\altaffiltext{4}{Institute of Astronomy, The Universe of Tokyo, \\
                 2-21-1 Osawa, Mitaka, Tokyo 181-0015}
\altaffiltext{5}{Kashima Space Research Center, National Institute of Information and Communications Technology,\\
                 893-1 Hirai, Kashima, Ibaraki 314-8510}
\email{takumi.nagayama@nao.ac.jp}


\maketitle


\begin{abstract}
We surveyed the (1,1), (2,2), and (3,3) lines of NH$_3$ and 
the H$_2$O maser toward the molecular cloud L935 
in the extended H\emissiontype{II} region NGC 7000
with an angular resolution of \timeform{1.6'} 
using the Kashima 34-m telescope.
We found five clumps in the NH$_3$ emission with a size of 
0.2--1 pc and mass of 9--452 \MO.
The molecular gas in 
these clumps has 
a similar gas kinetic temperature of 11--15 K 
and a line width of 1--2 km s$^{-1}$.
However, they have different star formation activities
such as the concentration of T-Tauri type stars and 
the association of H$_2$O maser sources.
We found that these star formation activities 
are related to the geometry of the H\emissiontype{II} region.
The clump associated with the T-Tauri type star cluster
has a high star formation efficiency of 36--62\%.
This clump is located near 
the boundary of the H\emissiontype{II} region and molecular cloud.
Therefore, we suggest that the star formation efficiency increases 
because of the triggered star formation.
\end{abstract}


\section{Introduction}

The star formation efficiencies (SFEs) of molecular clouds in the Milky Way Galaxy,
are typically observed to be 10\%.
The SFEs in nearby molecular clouds are $\simeq$ 3--15\% 
(\cite{swift2008}; \cite{evans2009}).
The observations of giant molecular clouds in the inner Galaxy indicate that
the SFEs in these clouds are of the order of a few percent \citep{myers1986}.
However, \citet{lada1992} found that 
three of five massive cores, NGC 2024, NGC 2068, and 
NGC 2071 exhibit higher SFEs of $\simeq$ 30--40\%.
It remains unclear why these SFEs are high.
\citet{lada1992} suggested that
the high gas densities and high gas mass may be 
required for the high SFE but there should be additional conditions,
because the other two cores exhibit low SFEs of $\simeq 7$\%.
However, additional conditions for high SFE are unknown,
and more observational investigations are required.
\renewcommand{\thefootnote}{\fnsymbol{footnote}}
\footnote[1]{Present address: Graduate School of Science and Engineering, 
Kagoshima University, 1-21-35 Korimoto, Kagoshima, Kagoshima 890-0065.}

\begin{figure*}[h]
\begin{center}
\FigureFile(80mm,80mm){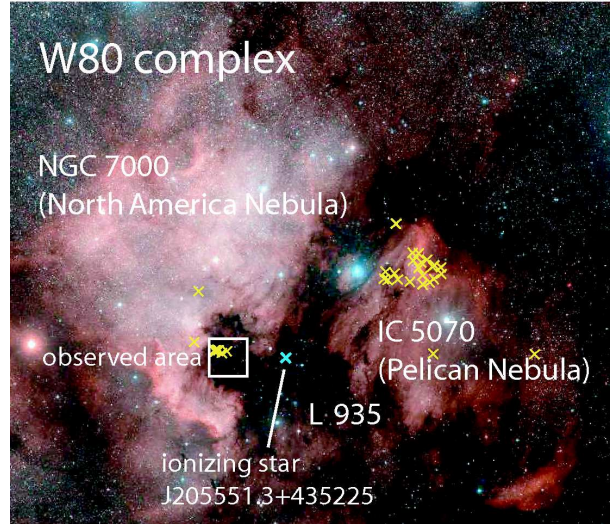}
\end{center}
\caption{Optical image of the W80 region (CalTech/Palomar)
that comprises two H\emissiontype{II} regions 
of NGC 7000 (North Amrica nebula) 
and IC 5070 (Pelican nebula) 
and dark lanes of L935.
Yellow crosses show the positions of T-Tauri type stars \citep{herbig1958}.
A blue cross shows the position of an O5 V type star \citep{comeron2005}.
}
\label{fig:1}
\end{figure*}

NGC 7000 is an extended H\emissiontype{II} region in the Cygnus X region.
On its southeastern side is a molecular cloud L935.
The $^{13}$CO emission in this molecular cloud is 
the brightest in the Cygnus X region \citep{dobashi1994}.
Figure \ref{fig:1} shows the optical image of NGC 7000 and L935.
Seven T-Tauri stars are clearly clustered at the boundary of the H\emissiontype{II} region 
\citep{herbig1958}.
This suggests that the star formation is triggered by 
the interaction of the H\emissiontype{II} region with the dense molecular gas.
A number of studies of
star formation in a cloud associated with an H\emissiontype{II} region
have been performed
(\cite{sugitani1989}; \cite{sugitani1991}; \cite{sugitani1994}; \cite{dobashi2001} 
\cite{deharveng2003}; \cite{deharveng2005}).
For example, in the nearby H\emissiontype{II} region, IC 5070, 
a molecular shell with an expanding velocity of 5 km s$^{-1}$, 
is found in the $^{12}$CO ($J$=1-0) line \citep{bally1980}.
They suggest that the T-Tauri type stars in IC 5070 are formed by the expanding shell.

Our aim is to investigate the relationship 
between dense molecular gas and star formation based on the SFE.
NGC700 has the advantage of allowing us to estimate the SFE because
T-Tauri type stars are associated with it and we can estimate the stellar mass accurately.
Therefore, we made the observations in the NH$_3$ line to estimate the mass of dense molecular gas.
We also surveyed an H$_2$O maser source 
that is associated with outflow from a young stellar object (YSO).
We adopted the distance to NGC 7000 to be 600 pc \citep{laugalys2002}.


\section{Observations}

\subsection{NH$_3$ Observations}

We observed NGC 7000 in the NH$_3$ lines with the Kashima 34-m telescope 
of the National institute of Information and Communications Technology (NiCT) from April 2007 to October 2008.
We made simultaneous observations in three inversion transitions of 
the NH$_3$ $(J,K)$ = (1,1), (2,2), and (3,3) lines 
at 23.694495, 23.722633, and 23.870129 GHz, respectively.
At 23 GHz, the telescope beam size was \timeform{1.6'} 
and the main beam efficiency ($\eta_{{\rm MB}}$) was 0.50.
We used a $K$-band HEMT amplifier whose system noise temperature was 150--250 K.
The relative pointing error was better than \timeform{0.2'}; 
this was checked by the observations of several H$_2$O maser sources at 22.235080 GHz.
All spectra were obtained with an 8192-channel FX-type 
spectrometer developed at Kagoshima University and NiCT.
Its bandwidth and frequency resolution are 256 MHz and 31.25 kHz, respectively.
The corresponding velocity coverage and velocity resolutions are 3200 km s$^{-1}$ 
and 0.39 km s$^{-1}$ at the NH$_3$ frequencies, respectively.
The total number of the observed positions is 311
and  the surveyed area is approximately \timeform{38'} $\times$ \timeform{11'} or 6.6 pc $\times$ 1.9 pc.
The NH$_3$ profiles were obtained at \timeform{1'} grid points of the equatorial coordinates.
All data were obtained with the position switch between the target and a reference position.
The reference position is 
$(\alpha, \delta)_{\rm (J2000)} = (\timeform{20h58m02.1s}, \timeform{+44D04'24"})$,
where no NH$_3$ emission was detected.
We integrated at least 20 min at each point.
The rms noise level was typically 0.20 K in the unit of the main beam brightness temperature defined by  
$T_{{\rm MB}} \equiv T_{{\rm A}}^{*} / \eta_{{\rm MB}}$, 
where $T_{{\rm A}}^{*}$ is the antenna temperature 
calibrated by the chopper wheel method \citep{kutner1981}.

Data reduction was performed using the UltraSTAR
package developed by the radio astronomy group at the University of Tokyo \citep{nakajima2007}.
In this paper, the intensities are presented in the main beam temperature.


\subsection{H$_2$O Maser Observations}

The single-dish observations of the 6$_{16} \rightarrow 5_{23}$ transition of the H$_{2}$O maser 
at 22.235080 GHz were made at the positions of the NH$_3$ emission peaks.
The observations were made with the Kashima 34-m telescope and the VLBI Exploration of Radio Astrometry (VERA) 
Iriki 20-m telescope.
In the observations made with the Kashima 34-m telescope,
the conversion factor from the antenna temperature to the flux density was 8 Jy K$^{-1}$ for a point source. 
The rms noise level of the H$_2$O maser spectra was less than 1 Jy after 
integration of 10--30 min.
The velocity resolution of the spectrometer was 0.42 km s$^{-1}$ at the H$_2$O maser frequency.
In the observations made with the VERA Iriki 20-m telescope,
the conversion factor from the antenna temperature to the flux density was 20 Jy K$^{-1}$. 
The rms noise level of the H$_2$O maser spectra was less than 1 Jy after 
integration of 30 min.
The velocity resolution of the spectrometer was 0.21 km s$^{-1}$ at the H$_2$O maser frequency.

The VLBI observations of the 6$_{16} \rightarrow 5_{23}$ transition of the H$_{2}$O maser were 
made with the VERA of the National Astronomical Observatory of Japan (NAOJ) on February 15, 2008.
The data were recorded using the VSOP terminal at a rate of 128 Mbit s$^{-1}$.
The recorded signals were correlated using the Mitaka FX correlator.
The spectral resolution was set at 31.25 kHz, 
which corresponds to a velocity resolution of 0.42 km s$^{-1}$.
The system noise temperature is approximately 140 K.
The phase calibrator was BL Lac (ICRF J220243.2+421639).
The phase-tracking center of the array was set at 
$\alpha$(J2000) = \timeform{20h57m56.717s}, 
$\delta$(J2000) = \timeform{+43D53'39.60"}.
We calibrated the data in the standard reduction procedure 
with the Astronomical Image Processing System (AIPS) of the National Radio Astronomy Observatory (NRAO).
The resultant rms noise level and synthesized beam size are $\simeq$ 1 Jy beam$^{-1}$ 
and $\simeq$ 1.4 $\times$ 0.7 mas with a position angle of \timeform{-46D}.


\section{Results}

\subsection{Distribution of NH$_3$ Clumps}
\label{sec:3.1}

\begin{figure*}[h]
\begin{center}
\FigureFile(160mm,80mm){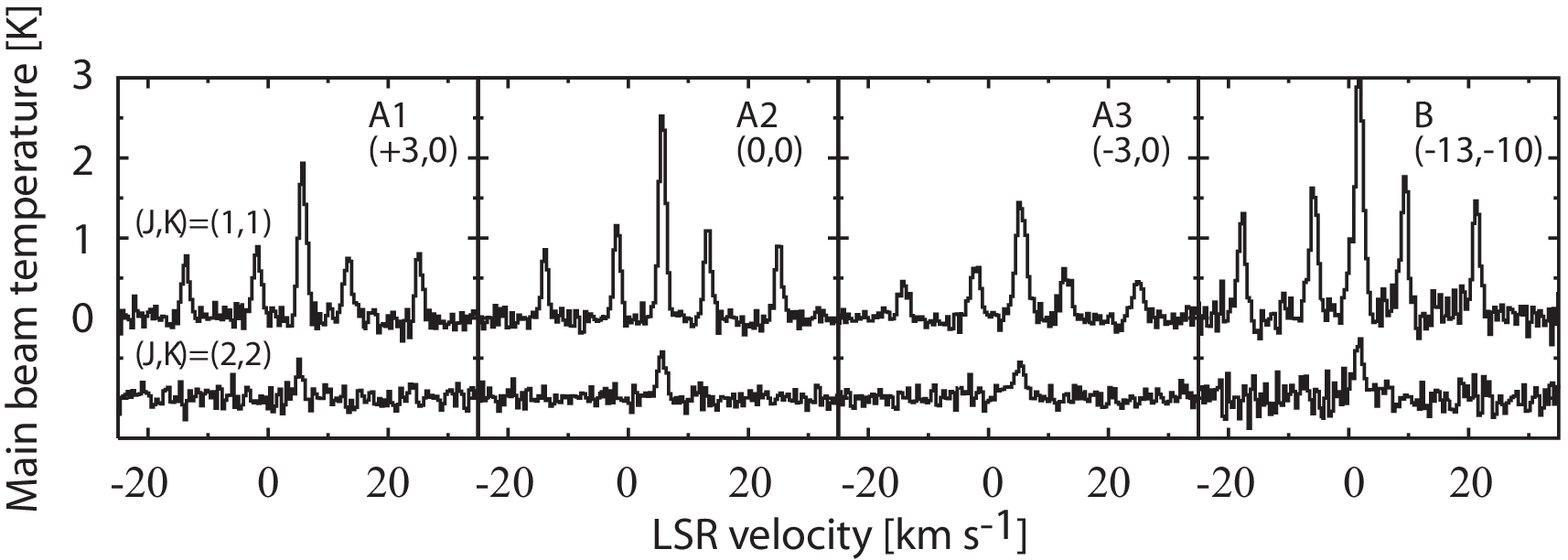}
\end{center}
\caption{Spectra in the NH$_3$ ($J,K$)=(1,1) and (2,2) lines 
observed at the peak position of the subclumps and the clump-B.
The position offsets in arcmin from 
$(\alpha, \delta)_{\rm (J2000)} = (\timeform{20h58m02.1s}, \timeform{+43D53'24"})$
are shown in the top-right corner.}
\label{fig:2}
\end{figure*}

We observed 311 positions in the survey.
NH$_3$ (1,1) and (2,2) lines were detected with a signal-to-noise ratio greater than
3 at 138 and 32 positions, respectively.
The (3,3) line was not detected at any positions in the observed area.
We show the profiles towards emission peaks in Figure \ref{fig:2}.

\begin{figure*}[h]
\begin{center}
\FigureFile(160mm,160mm){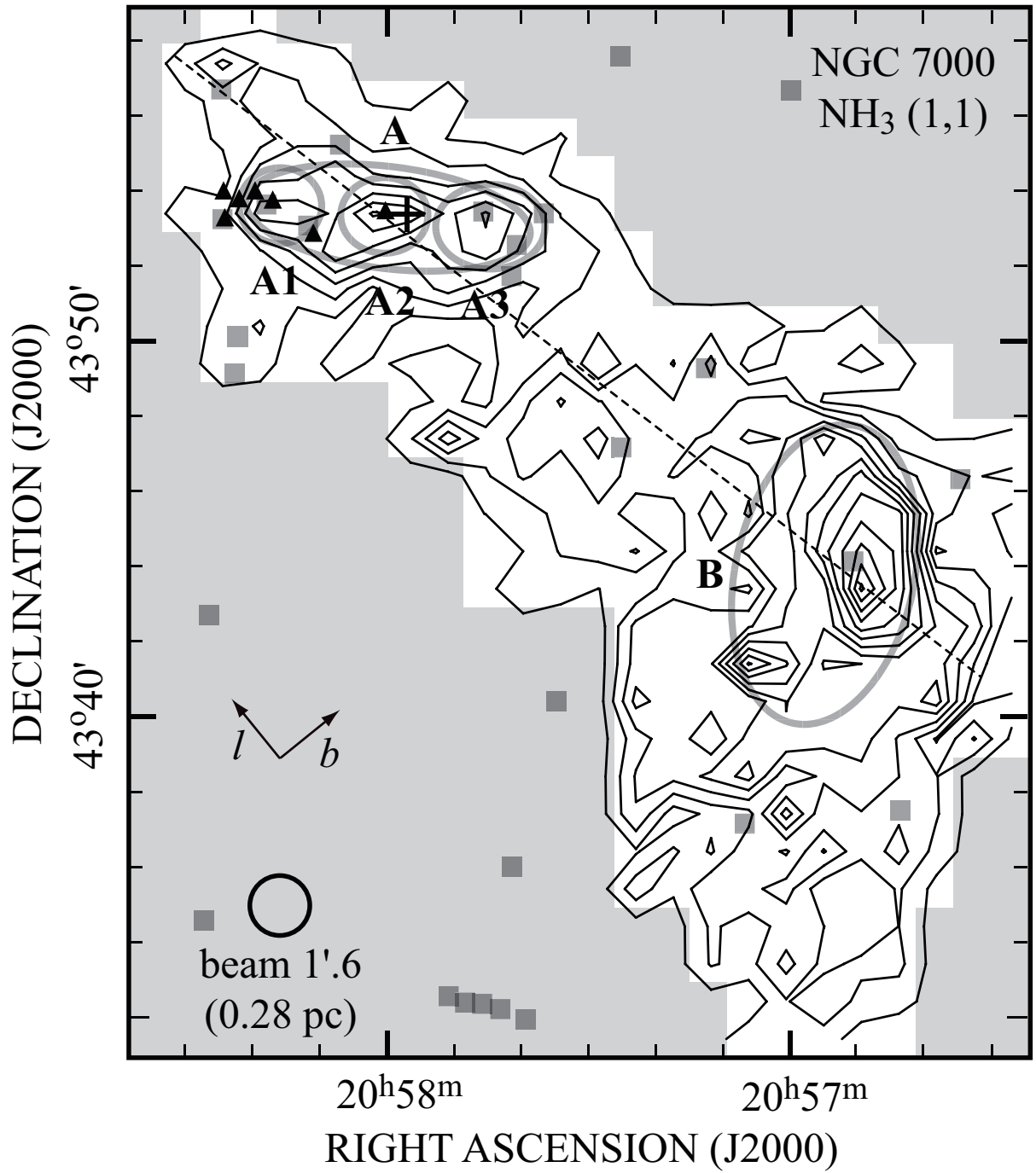}
\end{center}
\caption{Integrated intensity map of the main hyperfine component of the NH$_3$ (1,1) line.
The lowest contour and the contour interval are 0.3 and 0.6 K km s$^{-1}$, respectively.
The circle in the bottom-left corner shows the beam size.
The gray ellipses show the extents of the clumps and the subclumps.
The triangles and squares show the T-Tauri stars and MSX sources, respectively.
The cross shows the position of the H$_2$O maser.
The dashed line shows the axis of the position velocity map shown in Figure \ref{fig:5}.}
\label{fig:3}
\end{figure*}

Figure \ref{fig:3} shows the velocity integrated map of the main hyperfine component of 
the NH$_3$ (1,1) line over $v_{\rm LSR} = -3$ to 8 km s$^{-1}$.
The NH$_3$ (1,1) emission extends over 
\timeform{38'} $\times$ \timeform{11'} or 6.6 pc $\times $ 1.9 pc.
The distribution is elongated in the northeast-southwest direction.

\begin{figure*}[h]
\begin{center}
\FigureFile(160mm,160mm){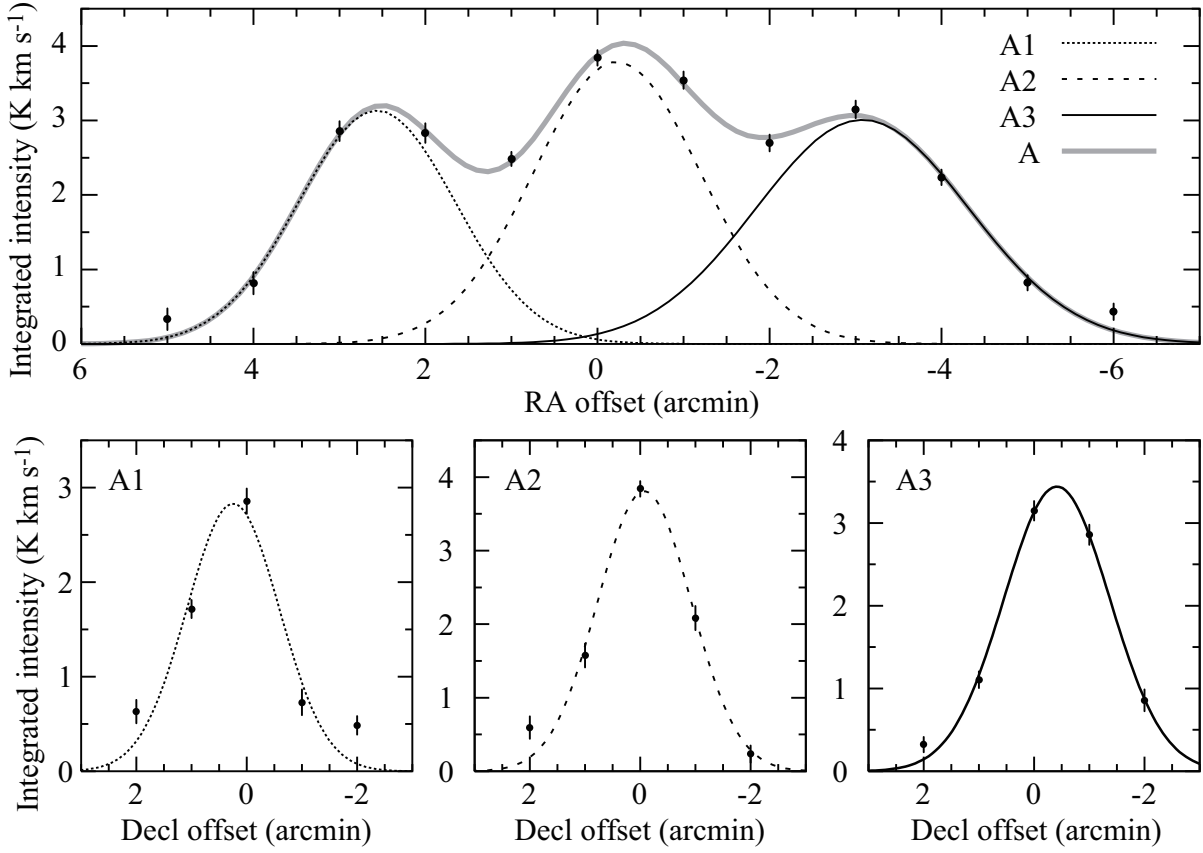}
\end{center}
\caption{The results of the Gaussian fitting to the intensities along the
right ascension and declination for subclump A1--A3.
The position offsets in arcmin from 
$(\alpha, \delta)_{\rm (J2000)} = (\timeform{20h58m02.1s}, \timeform{+43D53'24"})$ 
are shown.}
\label{fig:15}
\end{figure*}

We define a $clump$ as an isolated feature in the (1,1) integrated intensity distribution
with intensities stronger than the 3$\sigma$ noise level ($\gtrsim 0.6$ K km s$^{-1}$) 
both in the (1,1) and the (2,2) lines.
Based on this definition,
we identify two NH$_3$ clumps; 
we call them clump-A at the northeast and clump-B at the southwest.
We found local peaks in these clumps.
In clump-A, there are YSOs 
that may have been formed near three local peaks (see subsection 4.1).
In order to investigate the star formation of clump-A in detail,
we define three $subclumps$ as three local peaks in clump-A.
We call them subclump-A1, subclump-A2, and subclump-A3 from the east to the west.
The parameters of the clumps and subclumps are summarized in Table \ref{tab:1}.
As shown in Figure \ref{fig:3}, 
the shape of these subclumps appears to be an ellipse 
with the major axis along the right ascension 
and the minor axis along the declination.
Therefore, we estimate the subclump sizes of the major and minor axes 
using the FWHM of the Gaussian fitting to the intensities 
along the right ascension and declination, respectively.
Figure \ref{fig:15} shows the results of the Gaussian fitting.
This size is used to estimate the mass of each subclump.

\begin{figure*}[h]
\begin{center}
\FigureFile(160mm,160mm){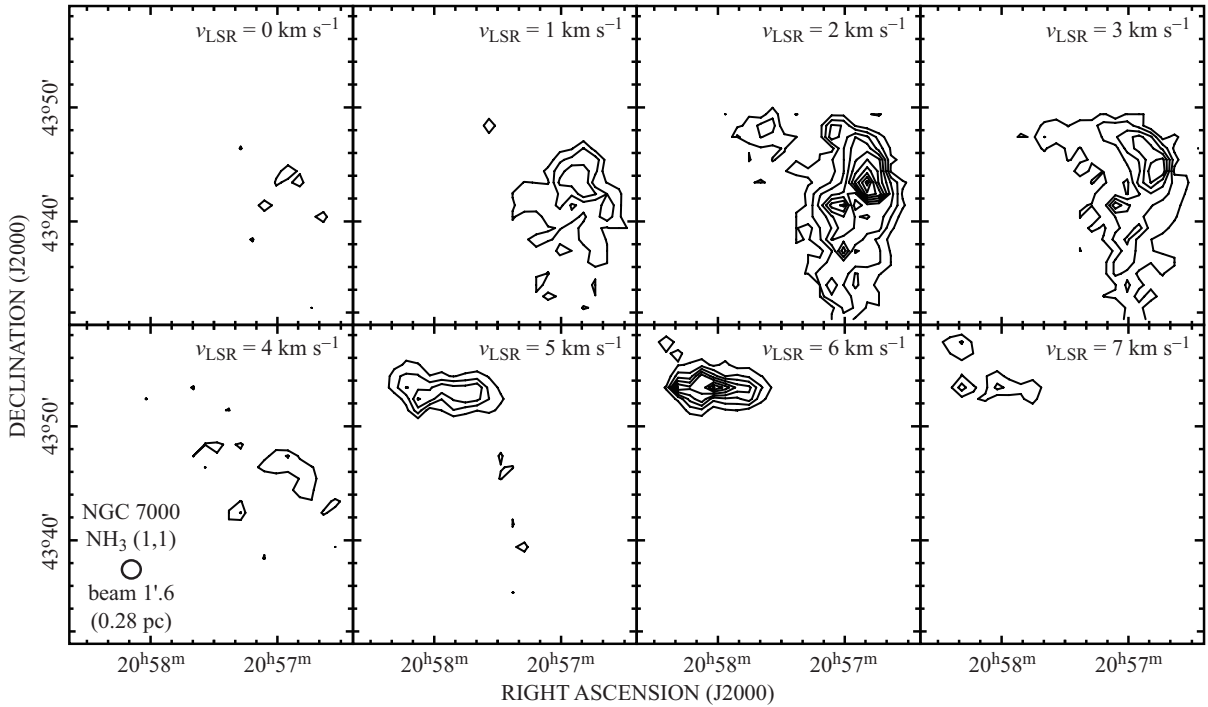}
\end{center}
\caption{Velocity channel maps in the NH$_3$ ($J,K$)=(1,1) line.
The lowest contour and the contour interval are 0.3 and 0.3 K, respectively.
The intensity is averaged in the velocity span of 1 km s$^{-1}$.}
\label{fig:4}
\end{figure*}

Clump-A extends over \timeform{11'} $\times$ 
\timeform{10'} or 1.9 pc $\times $ 1.8 pc (Figure \ref{fig:3}).
The velocity channel maps from $v_{\rm{LSR}} = 0$ 
to 8 km s$^{-1}$ are shown in Figure \ref{fig:4}.  
Clump-A appears at $v_{\rm{LSR}}$= 5--7 km s$^{-1}$ 
and clump-B appears at $v_{\rm{LSR}}$= 1--3 km s$^{-1}$.

\begin{figure*}[h]
\begin{center}
\FigureFile(80mm,80mm){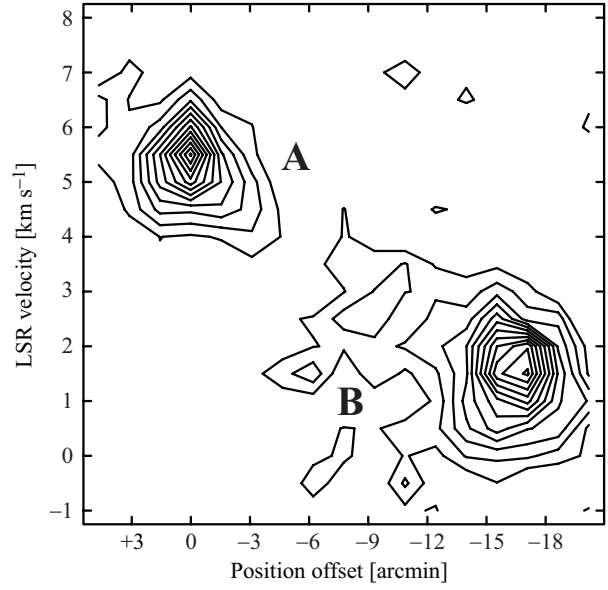}
\end{center}
\caption{Position-velocity map of the whole cloud in the NH$_3$ ($J,K$)=(1,1) line.
The lowest contour and the contour interval are 0.2 and 0.2 K at the $T_{{\rm MB}}$ unit, respectively.
The position offset is relative to the position at 
$(\alpha, \delta)_{\rm (J2000)} = (\timeform{20h58m02.1s}, \timeform{+43D53'24"})$ 
along the dashed line shown in Figure \ref{fig:3}.}
\label{fig:5}
\end{figure*}

Figure \ref{fig:5} shows the position-velocity diagram 
along the dashed line shown in Figure \ref{fig:3}.
Clump-A and clump-B are clearly separated in the position-velocity space.
It suggests that these clumps are not parts of a single object.
However, 
clump-A, clump-B, and other weak features between the clumps are aligned on the position-velocity diagram.
This suggests that these clumps may be formed from a single system (see subsection 4.3).

\begin{figure*}[h]
\begin{center}
\FigureFile(160mm,160mm){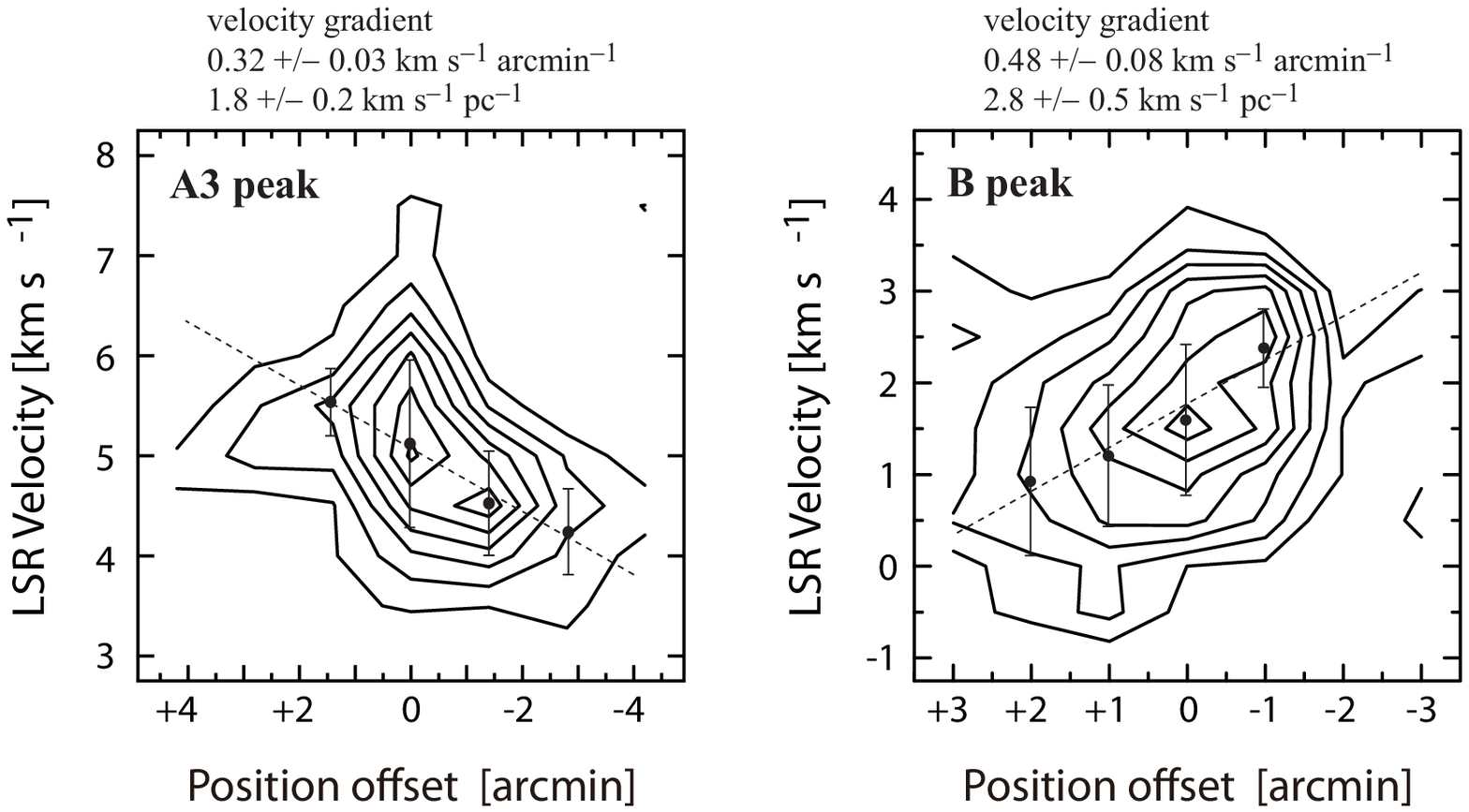}
\end{center}
\caption{Position-velocity maps of the NH$_3$ ($J,K$)=(1,1) 
line toward subclump-A3 (left) and clump-B (right).
The lowest contour and the contour interval are 0.3 and 0.3 K, respectively.
The position offset is relative to the peak positions at 
$(\alpha, \delta)_{\rm (J2000)} =$ (\timeform{20h57m45.5s}, \timeform{+43D53'24"})
along the position angle of \timeform{45D} for the A3 peak and 
(\timeform{20h56m49.9s}, \timeform{+43D43'24"}) 
along the position angle of \timeform{90D} for the B peak.}
\label{fig:6}
\end{figure*}

A velocity gradient is found in subclump-A3.
Figure \ref{fig:6} shows the position-velocity diagram of subclump-A3.
The estimated velocity gradient is 1.66 $\pm$ 0.57 km s$^{-1}$ pc$^{-1}$.
The linewidth of the peak spectrum is 2.11 $\pm$ 0.08 km s$^{-1}$, 
and it is 1.4 times broader than the other spectra in clump-A.
The other subclumps do not show the velocity gradient.

Clump-B extends over \timeform{15'} $\times$ \timeform{19'} or 
2.6 pc $\times $ 3.3 pc (Figure \ref{fig:3}).
In the velocity channel maps, 
clump-B appears at $v_{\rm{LSR}}$= 1--4 km s$^{-1}$ (Figure \ref{fig:4}).
A velocity gradient is found in the emission peak of clump-B.
Figure \ref{fig:6} shows the position-velocity diagram of the peak of clump-B.
The estimated velocity gradient is $-1.26 \pm 0.23$ km s$^{-1}$ pc$^{-1}$.

The $^{13}$CO ($J$=1-0) emission was detected by \citet{dobashi1994} 
with a velocity range of $v_{\rm{LSR}}$= 1.5--7.5 km s$^{-1}$
at the mid position of clump-A and clump-B.
It comprises two velocity components of clump-A and clump-B.
The spatial distribution in the NH$_3$ line is similar to that in $^{13}$CO line,
although the $^{13}$CO observations were made
with a lower angular resolution (\timeform{2.7'}) and sparsed sampling (\timeform{5'}).
Clump-B is located at the CO peak both on the sky and in the velocity.
Clump-A is located on the northeast side of the CO cloud; 
it faces the H\emissiontype{II} region.


\subsection{Physical Parameters of Clumps and Subclumps}
\label{sec:3.2}

We estimated the physical parameters of the clumps and the subclumps.
NH$_3$ is a very well-studied molecule
with which to investigate the physical conditions of the dense molecular gas. 

\begin{figure*}[h]
\begin{center}
\FigureFile(160mm,100mm){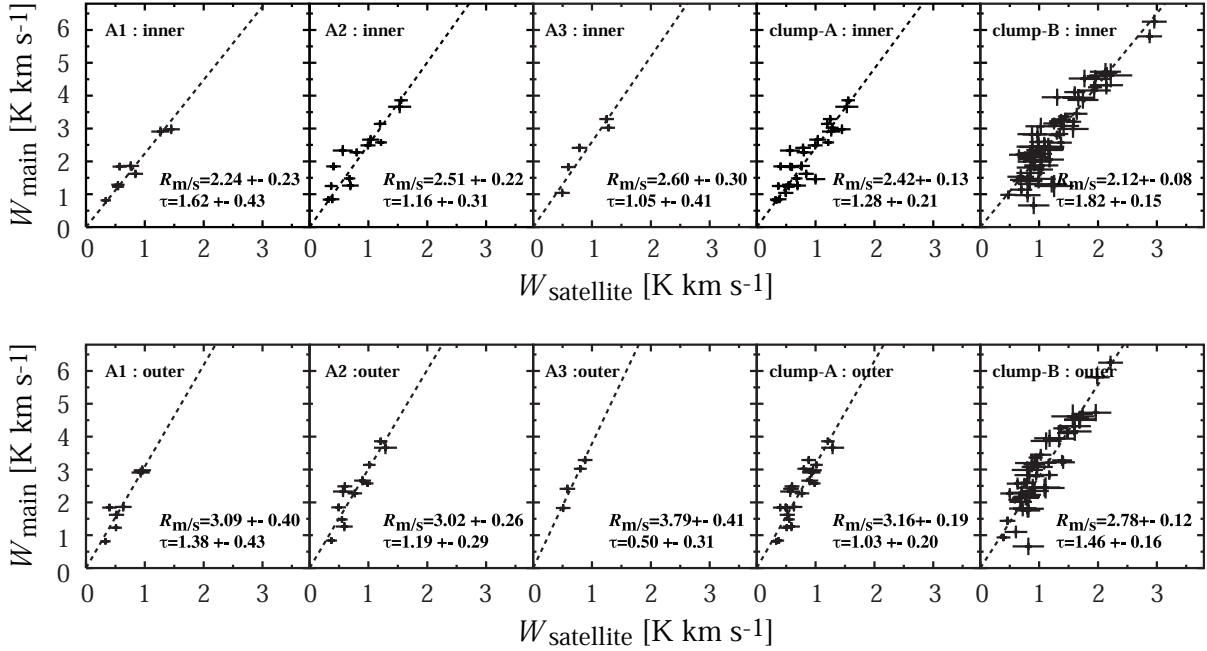}
\end{center}
\caption{Correlations of the integrated intensity in the (1,1) main and satellite lines.
The correlations of the main to the inner and outer satellite lines
are shown in the top and bottom panel, respectively.
The data detected over the 3 $\sigma$ level in both the main and the satellite lines are plotted.
The error bar shows the rms noise (1$\sigma$).
The estimated optical depth is shown in the bottom of each panel.}
\label{fig:7}
\end{figure*}

The NH$_3$ lines are split by the quadrupole hyperfine interaction.
Optical depths can be directly determined 
from the intensity ratio of the main to the satellite lines.
Because we detect the hyperfine structure in the (1,1) line, 
the optical depth, $\tau_{(1,1)}$, can be derived from the intensity ratio.
Figure \ref{fig:7} shows the correlation of the integrated intensities 
of the main and the satellite lines.
We estimated two intensity ratios of the inner and outer satellite lines to the main line.
The optical depths estimated from these two ratios are the same within the error.
We list the optical depths of the clumps and the subclumps in Table \ref{tab:2},
and these are found to be in the range of 0.8--1.6.

\begin{figure*}[h]
\begin{center}
\FigureFile(160mm,100mm){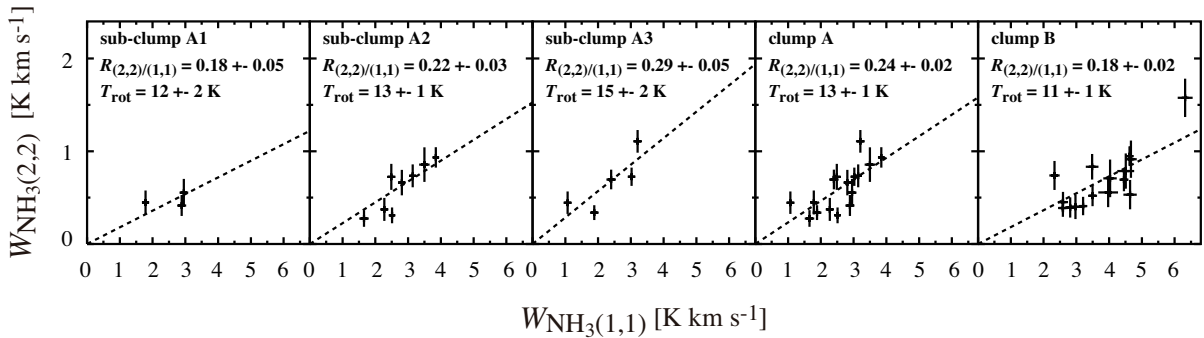}
\end{center}
\caption{Correlations of the integrated intensity in the (1,1) and (2,2) lines.
The data detected over the 3 $\sigma$ level in both the (1,1) and the (2,2) lines are plotted.
The error bar shows the rms noise (1$\sigma$).
The estimated $T_{\rm rot}$ is shown in each panel.}
\label{fig:8}
\end{figure*}

We estimated the NH$_3$ rotational temperature from the intensity ratio of 
the (2,2) line to the (1,1) line using the method shown by \citet{ho1983}.
Figure \ref{fig:8} shows the correlations of the integrated intensities in the (1,1) and (2,2) lines. 
The rotational temperatures of all clumps and subclumps are 11--15 K and the same within the error.
Using the collisional excitation model (\cite{walmsley1983}; \cite{danby1988}),
the rotational temperature is estimated to be very close to the gas kinetic temperature, $T_{\rm kin}$, 
for $T_{\rm rot} < 15$ K.
Therefore, $T_{\rm kin}$ should be approximately 13 K.
The estimated temperatures are listed in Table \ref{tab:2}.

We derive the total column density of NH$_3$, $N({\rm NH_3}$), 
from the column density in the (1,1) line, 
assuming the local thermodynamic equilibrium (LTE) condition 
for the molecules in the clumps \citep{rohlfs1996}.
The estimated total column densities of the clumps are listed in Table \ref{tab:2}
and they are in the range of  
$N({\rm NH_3})$=(1.8--3.7) $\times$ 10$^{15}$ cm$^{-2}$.

We derived the molecular gas mass for each of the clumps and subclumps 
using two methods.
One is the LTE mass 
that is derived from the deconvolved size 
and the column density with the assumed abundance ratio.
Using a model of a uniform density share with 40\% helium in mass,
the LTE mass is given by
\begin{eqnarray}
M_{\rm LTE}
=
467
\left(
\frac{R}{[{\rm pc}] }
\right)^2
\left(
\frac{N({\rm NH}_3)}{10^{15} [{\rm cm}^{-2}]}
\right)
\left(
\frac{X({\rm NH}_3)}{10^{-7}}
\right)^{-1}
\MO,
\end{eqnarray}
where $R$ is the radius of the sphere, and 
$X({\rm NH}_3)$ is the abundance of NH$_3$ relative to H$_2$.
For clump-A and B, $R$ is given from a geometrical mean of the 
major and minor axes of an apparent ellipse after beam deconvolution.
For subclumps A1--A3, $R$ is given from a geometrical mean of FWHMs of 
the Gaussian fitting along the right ascension and declination. 
The estimated sizes in diameter are in the range of 0.6--1 pc 
for the clumps and 0.2--0.3 pc for the subclumps. 

\citet{ho1983} reviewed that
the abundance of NH$_3$ relative to H$_2$ has been estimated to 
range from $10^{-7}$ in the core of the dark cloud L183 \citep{ungerects1980} 
up to $10^{-5}$ in the hot core of the Orion KL \citep{genzel1982} and
the ion-molecule chemistry produces 
an abundance of the order of $10^{-8}$ \citep{prasad1980}.
We use the abundance ratio of $10^{-7}$,
because the estimated kinetic temperature and the clump size in NGC 7000
are close to those of L183.
Using this abundance,
the hydrogen column density derived from NH$_3$ is derived to be
$N$(H$_2$)=(1.8--3.7) $\times$ 10$^{22}$ cm$^{-2}$.
This corresponds to $A_{V} \simeq$ 10--20 mag by assuming the conversion factor from the relation 
$N({\rm H_2})/A_{V}=1.87 \times 10^{21}$ atoms cm$^{-2}$ mag$^{-1}$ \citep{bohlin1978}.
\citet{comeron2005} estimated the extinction of the molecular cloud 
to be $A_{V} \simeq$10--30 mag using 2 MASS archive data.
These two values of $A_V$ are consistent within a factor.
Moreover, the hydrogen column density estimated from $^{13}$CO \citep{dobashi1994}
is consistent with our estimation within a factor of 2--3,
although the observation grid is different. 
The LTE masses of clump-A  and clump-B are estimated to be 
95 and 452 \MO, respectively.
The LTE masses of subclump-A1, subclump-A2, and subclump-A3 
are estimated to be 12, 20, and 9 \MO, respectively.

The other mass estimation is the virial mass, $M_{\rm vir}$.
This is calculated as $M_{\rm vir} = k R \Delta v^2 \MO$,
where $k$ is taken as 210 based on a uniform density sphere
(\cite{maclaren1988}), $R$ is the radius
of the clump, and $\Delta v$ is the half-power line width.
The virial masses of clump-A and clump-B are estimated to be 125 and 350 \MO, 
respectively.
The virial masses of subclump-A1, subclump-A2, and subclump-A3
are not estimated, because their sizes are small.

We found that the LTE mass and the virial mass are consistent by a factor of 1.3.
From the above mentioned comparisons such as $A_V$, $^{13}$CO, and derived mass, 
we could estimate the actual masses of molecular gas within a factor of 2--3. 
All these derived parameters are summarized in Table \ref{tab:2}.


\subsection{H$_{2}$O Maser Source in Subclump-A2}
\label{sec:3.3}

We conducted an H$_2$O maser survey of the NH$_3$ clumps and subclumps.
We discovered a new H$_2$O maser emission at the NH$_3$ peak in subclump-A2.
However, no H$_2$O maser emission was detected in the others with the upper limit of 3 Jy (Figure \ref{fig:3}).

\begin{figure*}[h]
\begin{center}
\FigureFile(80mm,80mm){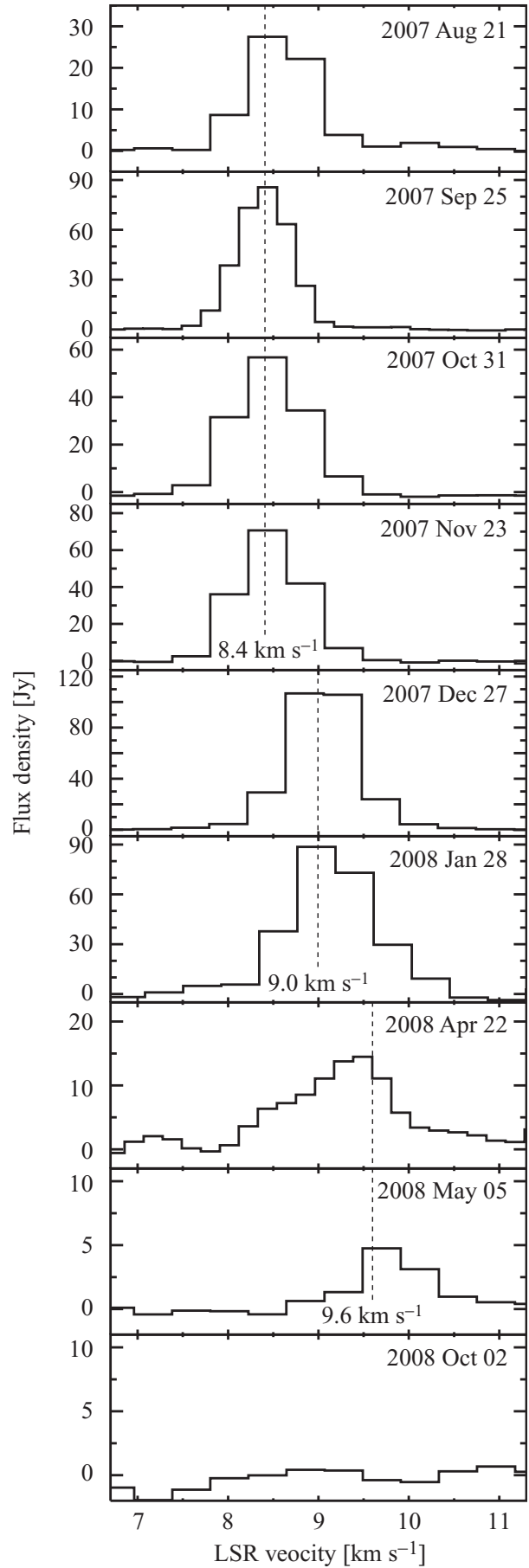}
\end{center}
\caption{H$_2$O maser spectra obtained by our single-dish monitoring observations.
          Two velocity jumps are found between November and December of 2007,
          and between January and April of 2008.
          The maser emission disappeared in October 2008.}
\label{fig:9}
\end{figure*}

We conducted monitoring observations of the maser source from August 2007 to May 2008.
The obtained spectra are shown in Figure \ref{fig:9}.
All spectra show a single velocity component with a narrow linewidth (FWHM) of $\simeq$ 1 km s$^{-1}$. 
This maser emission is time-variable between 107 Jy in December 2007 
and 5 Jy in May 2008.
We found that the maser emission disappeared with the upper limit of 2.1 Jy in October 2008.
The LSR velocity of the maser jumped from 8.4 to 9.0 km s$^{-1}$ 
during the period from November 2007 to December 2007,
and from 9.0 to 9.6 km s$^{-1}$ from January 2008 to April 2008. 
This means that
there were two maser spots and the lifetime of each spot is less than a year.

\citet{claussen1996} reported 
that an H$_2$O maser associated with a 
low-mass star is variable on a timescale ranging from months to a year.
The H$_2$O maser in subclump-A2 shows variations on the same timescale. 
Therefore, the H$_2$O maser should be associated with low-mass stars.

The detected H$_2$O maser of $v_{\rm{LSR}} \simeq$ 9 km s$^{-1}$ is 
redshifted with respect to the ambient gas velocity of $\simeq$ 5 km s$^{-1}$ 
observed in the NH$_3$ line. 
This suggests that the maser may be associated with outflows from the protostar.

To identify the counterpart of the maser, 
we should obtain an accurate position of the maser.
Therefore, we conducted VLBI observations on February 15, 2008,
to determine its position.
A single feature with a size of 1.6 mas $\times$ 0.7 mas or 0.96 $\times$ 0.42 AU was detected.
The position of this maser feature was obtained to be 
$(\alpha, \delta)_{\rm J2000} = (\timeform{20h57m57.01s}, \timeform{+43D53'28.5"})$
by a fringe rate analysis.
The position uncertainty was approximately \timeform{0.1"}.
However, we can find no visible star or optical feature suggesting an outflow 
at the position of the maser in the DSS2 images.
We discuss the counterpart of the maser in subsection 4.1.


\section{Discussion}

\subsection{Observed Star Formation Activity in Each Clump and Subclump}

To investigate the star formation in our identified NH$_3$ clumps,
we examined the distribution of young stars reported in the previous observations.
In clump-A, 
\citet{herbig1958} found seven emission-line stars, and
\citet{cohen1979} confirmed that these stars are T-Tauri type stars.
T-Tauri type stars are suitable for the mass estimation
because their mass can be well estimated using
the H-R diagram \citep{cohen1979}.
We estimate the mean and total masses of T-Tauri type stars in clump-A 
to be 1.3 and 9.0 \MO, respectively.
We consider that 
these T-Tauri type stars indicate the lower limit of the star formation activity.
We found 32 and 11 infrared sources listed in the 2MASS and MSX catalogues, respectively, in clump-A.
We consider that these infrared sources indicate the upper limit
of the star formation activity,
although some of these sources may be the fore/background sources.
There is a large difference in the distribution of 
these T-Tauri type stars and infrared sources in each subclump, 
suggesting differences in star formation activity.

We found that five of seven T-Tauri type stars are concentrated in subclump-A1.
The total mass of these T-Tauri stars is 6.9 \MO.
The concentration of the T-Tauri type stars suggests that
the star formation of subclump-A1 is the most active.
There are 15 2MASS sources and 3 MSX sources in subclump-A1.
The T-Tauri type stars and the majority of 2MASS sources are found
on the east side of subclump-A1,
where the H\emissiontype{II} region is located.

\begin{figure*}[h]
\begin{center}
\FigureFile(160mm,100mm){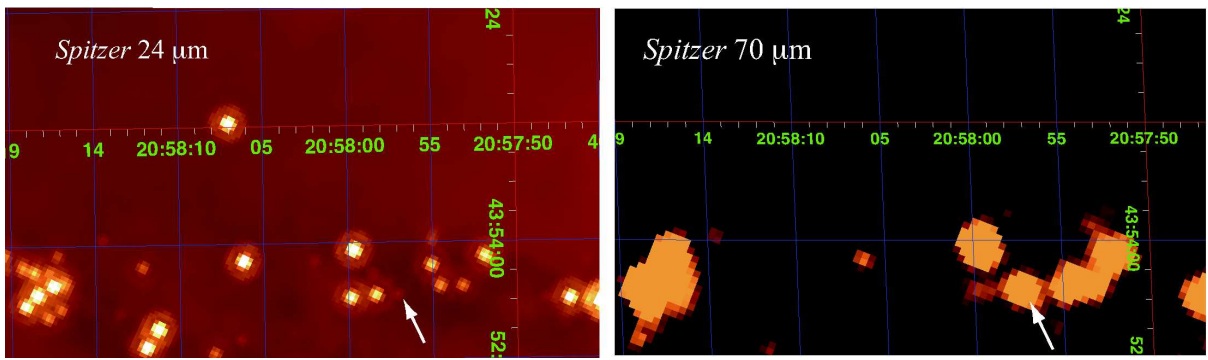}
\end{center}
\caption{$Spitzer$ images at 24 $\mu$m (left) and 70 $\mu$m (right).
A white arrow shows the position of the H$_2$O maser.}
\label{fig:10}
\end{figure*}

\begin{figure*}[h]
\begin{center}
\FigureFile(80mm,80mm){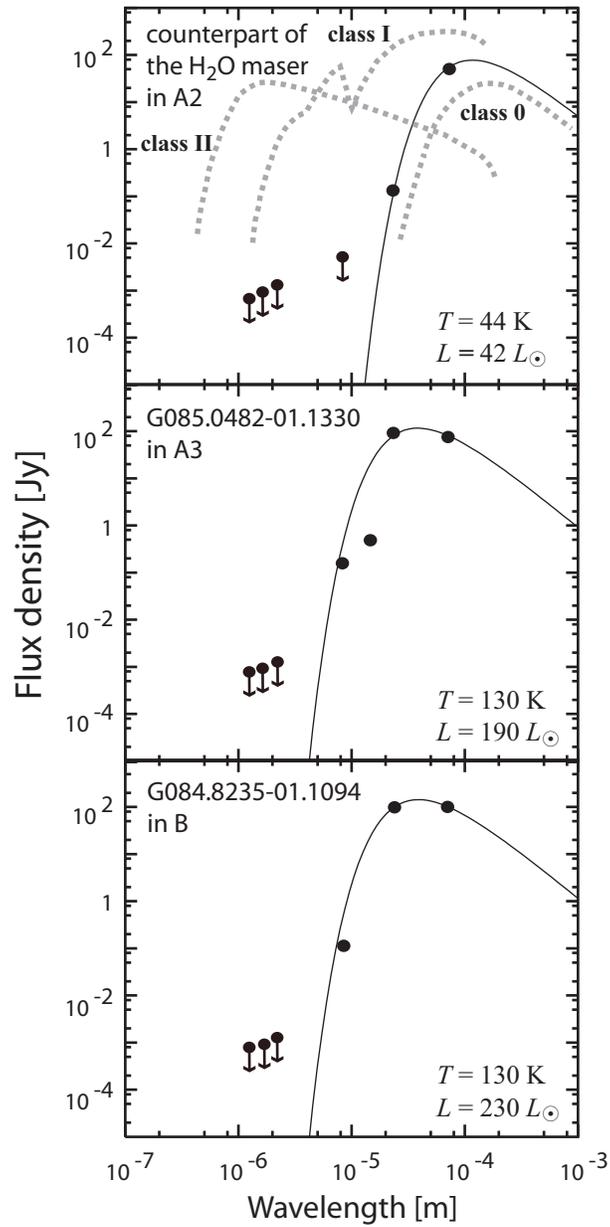}
\end{center}
\caption{Spectral energy distribution of the three infrared sources.
          The solid line shows the best fit blackbody with parameters
          shown at the right-bottom corner in each panel.
          The gray dashed lines in the top panel show typical spectra of 
          class 0, I, and II (\cite{bachiller1996}).}
\label{fig:11}
\end{figure*}

We found a new H$_2$O maser source in subclump-A2.
\citet{furuya2003} reported that $\simeq$ 40\% of class 0, 
$\simeq$ 4\% of class I, and no class II low-mass protostars emit the H$_2$O maser.
We found the counterpart of the H$_2$O maser in the $Spitzer$ infrared
images at 24 and 70 $\mu$m (Figure \ref{fig:10}).
The 70 $\mu$m emission of the counterpart is centered at 
($\alpha, \delta)_{\rm J2000} = (\timeform{20h57m57.33s}, \timeform{+43D53'27.9"})$
and extends over a 20 $\times$ \timeform{20"} (12000 $\times$ 12000 AU) area.
The position of the counterpart is consistent with 
that of the H$_2$O maser obtained by our VLBI observations.
The $Spitzer$ source should be a protostar associated with the H$_2$O maser.
We show a spectrum of the $Spitzer$ source at six wavelengths in Figure \ref{fig:11}.
At the wavelengths of 1.25--8.28$\mu$m,
the counterpart is not detected with 2MASS and MSX, and we show the upper limits.
A single-temperature blackbody radiation through the data points at 24 and 70 $\mu$m
is consistent with the upper limits.
Its bolometric luminosity is estimated to be 42 \LO.
Both the spectrum shape and the bolometric luminosity are consistent with 
those of a class 0 protostar \citep{bachiller1996}.
Therefore, the counterpart of the H$_2$O maser should be a class 0 protostar.
To confirm this, the submillimeter observations are required.

One T-Tauri type star and 5 2MASS sources are found in subclump-A2.
A nebulosity at
$(\alpha, \delta)_{\rm J2000} = (\timeform{20h57m55s}, \timeform{+43D53'40"})$ 
is visible in the DSS2 images in the $B$, $R$, and $I$-bands.
This is considered to be a reflection nebula, 
because it is continuously detected at the optical wavelength.

T-Tauri type stars are not found in subclump-A3.
Four 2MASS and four MSX sources are found.
These MSX sources would be the YSOs embedded in the dust envelope
because they are invisible in the DSS2 and 2MASS images.
G085.0482-01.1330 identified by MSX is located at the peak position of the (1,1) line in subclump-A3.
We found a velocity gradient at this position (see subsection \ref{sec:3.1}).
The velocity gradient would be due to the simple core rotation or
the outflow from G085.0482-01.1330. 
This source is located at the center of the velocity gradient.
The spectrum of this source, shown in Figure \ref{fig:11},
is similar to that of a class I protostar.
The bolometric luminosity was estimated to be 190 \LO.

T-Tauri type stars and H$_2$O maser sources are not found in clump-B.
We found 24 2MASS and 6 MSX sources.
One of the MSX sources identified as G084.8235-01.1094,
is located at the peak position in the (1,1) line.
The velocity gradient is found in the (1,1) line at this position.
The spectrum of this source, shown in Figure \ref{fig:11},
is similar to that of a class I protostar.
Its bolometric luminosity was estimated to be 230 \LO.

We consider that
the star formation of clump-A is more active than that of clump-B.
In clump-A,
both the number and the total mass of stars in subclump-A1 are 
larger than those of subclump-A2 and A3.


\subsection{Star Formation Efficiency}
\label{sec:4.2}

We have presented the stellar mass of each clump in the previous subsection.
In this subsection, we examine
the relation between the stellar and the molecular gas masses.
We estimate the star formation efficiency given by SFE=$M_{\rm star}/(M_{\rm star}+M_{\rm gas})$.

The SFE of subclump-A1 is estimated to be $\simeq$ 36\%
from the stellar mass of 6.9 \MO and gas mass of 12 \MO.
We use the stellar mass of the identified T-Tauri type stars in this estimation.
However, other T-Tauri type stars may be associated with the NH$_3$ clumps but located just behind them.
In this case, these T-Tauri type stars could be detected in the $K$-band of 2MASS.
We obtained the extinction in the $K$-band to be $A_K \simeq$ 1 mag
using $A_V \simeq$ 10 mag estimated from the NH$_3$ column density 
and the extinction law of $A_K$/$A_V$ = 0.112 \citep{rieke1985}.
For the identified T-Tauri type stars, 
the $K$-band magnitude of 2MASS is 8.5--11.5 mag.
Therefore, the $K$-band magnitude of a T-Tauri type star 
located behind the clumps is estimated to be 9.5--12.5 mag.
This value is brighter than the 2MASS $K$-band detection limit of 14.3 mag (SNR = 10).
We found ten 2MASS sources that are not identified as T-Tauri type stars 
in subclump-A1.
In the case that all of them are T-Tauri type stars with a mass of 1.3 \MO,
the SFE of subclump-A1 increases to $\simeq$ 69\%.

In subclump-A2, a T-Tauri type star with a mass of 0.8 \MO is found.
Therefore, 
its SFE is estimated to be $\simeq$ 4\% from the gas mass of 20 \MO.
The SFE averaged over the whole clump-A is estimated to be $\simeq$ 8\%
from the stellar mass of 7.7 \MO and gas mass of 95 \MO.
The SFEs of both subclump-A3 and clump-B might be 0\%
because no T-Tauri type star is found there.
In the case that the 2MASS sources are included in the stellar mass estimation,
the SFEs of subclump-A2, A3, and the whole of clump-A are estimated to be 23--36\%.
This value is close to the SFE of subclump-A1.
However, clump-B shows lower SFE of 6\% even in this case.

In either case, including only T-Tauri type stars or also 
the 2 MASS sources,
the SFE of subclump-A1 is estimated to be 36--62\%;
this is higher than the SFEs of the other molecular clouds.
In order to make a fair comparison, 
we revisited the SFEs of the following three molecular clouds using the same procedure.
The SFE of NGC 2264 is estimated to be 11\% from 
the stellar mass of the OB and T-Tauri type stars of $119 \MO$ (\cite{dahm2005})
and the molecular gas mass traced in the NH$_3$ line of $1000 \MO$ (\cite{lang1980}).
The SFE of NGC 1333 is estimated to be 16\% from 
the stellar mass of the T-Tauri type stars of $17 \MO$ (\cite{aspin2003})
and the molecular gas mass traced in the NH$_3$ line of $106 \MO$ (\cite{ladd1994}).
The SFE of L1228 is estimated to be 8\% from 
the stellar mass of the T-Tauri type stars of $1 \MO$ (\cite{kun2009})
and the molecular gas mass traced in the NH$_3$ line of $12 \MO$ (\cite{anglada1994}).
The SFE of subclump-A1 is close to $\simeq$ 42\% estimated at NGC 2024 and NGC 2068 
from the CS observations (\cite{lada1992}).
The derived SFEs of individual clumps are summarized in Table \ref{tab:3}.
The values of ``Total'' in Table \ref{tab:3} 
correspond to the upper limits of the SFEs
for the case in which the all 2MASS sources we found
are associated with the clump or the subclump.


\subsection{Geometry}

Although clump-A and clump-B are adjacent on the sky,
there is a big difference in the star formation activities. 
Because these clumps are close to NGC 7000,
the difference may be due to the H\emissiontype{II} region.
Therefore, we discuss the geometry of the clumps and the H\emissiontype{II} region.

The optical image (Figure \ref{fig:1}) shows that
the clumps are located in the foreground of the H\emissiontype{II} region.
Subclump-A1 is the nearest to and clump-B is the farthest from the H\emissiontype{II} region on the sky.

To investigate the three-dimensional structure of clump-A, 
we estimated the length along the line of sight, $l$, 
derived as $l=N({\rm H}_2) / n_{\rm cr}$, 
where $N({\rm H}_2)$ and $n_{\rm cr}$ are the hydrogen column density
and the critical density in the NH$_3$ line, respectively.
When we use $n_{\rm cr} \simeq$ 10$^{4}$ cm$^{-3}$ \citep{myers1983},
the lengths of subclump-A1, A2, and A3 are estimated to be 
$l \simeq$ 0.8, 0.6, and 0.6 pc, respectively.
These values are 2--3 times longer than the sizes on the sky.
The three subclumps may be the end-on view of the elephant trunks observed in M16.

As seen in subsection 3.2, clump-A and clump-B are spatially separated.
This suggests that these two clumps are gravitationally unbound.
The total LTE mass of the two clumps is 547 \MO.
If the two clumps are gravitationally bound,
the enclosed mass is estimated to be 11000 \MO from 
their separation of 2.9 pc and relative velocity of 4 km s$^{-1}$.
The LTE mass is $\sim$ 1/20 of the enclosed mass.
The mass of the cloud around the two clumps traced in the $^{13}$CO line 
is estimated to be $3400 \MO$ from the column density of $1.15 \times 10^{22}$ cm$^{-2}$
\citep{dobashi1994}.
Both the total mass of the two clumps and the $^{13}$CO cloud 
is smaller than the enclosed mass.
Therefore, clump-A and clump-B are gravitationally unbound.
However, clump-A, clump-B, and other weak features 
between the clumps are aligned on the position-velocity diagram (Figure \ref{fig:5}).
This suggests that they are also aligned in the three-dimensional structure.

\begin{figure*}[h]
\begin{center}
\FigureFile(80mm,80mm){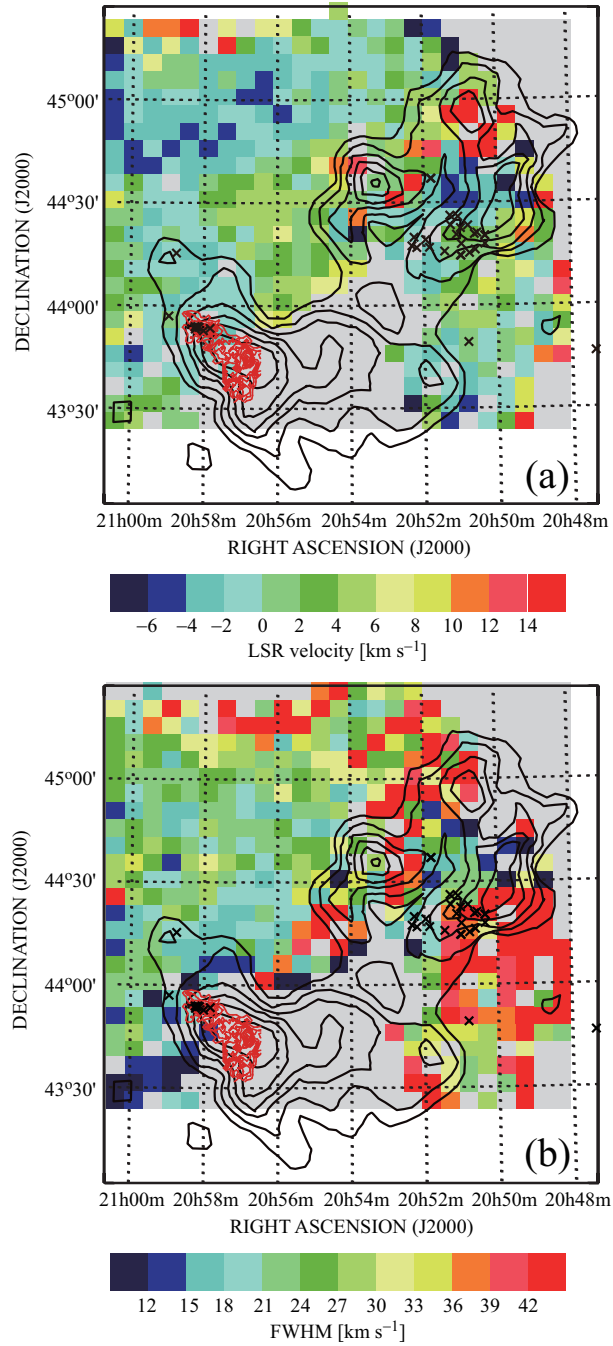}
\end{center}
\caption{(a) The LSR velocity map of the H\emissiontype{$\alpha$} emission (color; \cite{fountain1983})
on which the $^{13}$CO integrated intensity map (black contour; \cite{dobashi1994}) 
and the NH$_3$ (1,1) integrated intensity map (red contour) are superimposed.
The black cross is the emission line star \citep{herbig1958}.
(b) The FWHM map of the H\emissiontype{$\alpha$} emission (color; \cite{fountain1983})
superimposed on the same maps as that shown in (a).}
\label{fig:12}
\end{figure*}

We compare the LSR velocities of the H\emissiontype{II} region and the molecular gas.
Figure \ref{fig:12}(a) shows the LSR velocity map of the H\emissiontype{$\alpha$} 
emission \citep{fountain1983}
superimposed on the $^{13}$CO integrated intensity map \citep{dobashi1994}
and the NH$_3$ (1,1) integrated intensity map.
The LSR velocity of the H\emissiontype{$\alpha$} emission line is approximately 0 km s$^{-1}$ at the position overlapped 
with the NH$_3$ clumps, and 
4--5 km s$^{-1}$ on both the eastern and the western sides of the NH$_3$ clumps.
The LSR velocities of clump-A and clump-B are 5.5 and 1.5 km s$^{-1}$, respectively.
Clump-A is redshifted with respect to the H\emissiontype{II} region.
We show the FWHM map of the H\emissiontype{$\alpha$} emission \citep{fountain1983} in 
Figure \ref{fig:12}(b).
The FWHM of the H\emissiontype{$\alpha$} emission around the NH$_3$ clumps is 10--20 km s$^{-1}$.
This value is narrower than the FWHM at the other position.
These characteristics can be interpreted as indicating that 
the redshifted component of the H\emissiontype{$\alpha$} emission is blocked by clump-A,
and only ionized gas located in the foreground of clump-A would be observed.
Therefore, clump-A would be surrounded by the ionized gas.

\begin{figure*}[h]
\begin{center}
\FigureFile(120mm,100mm){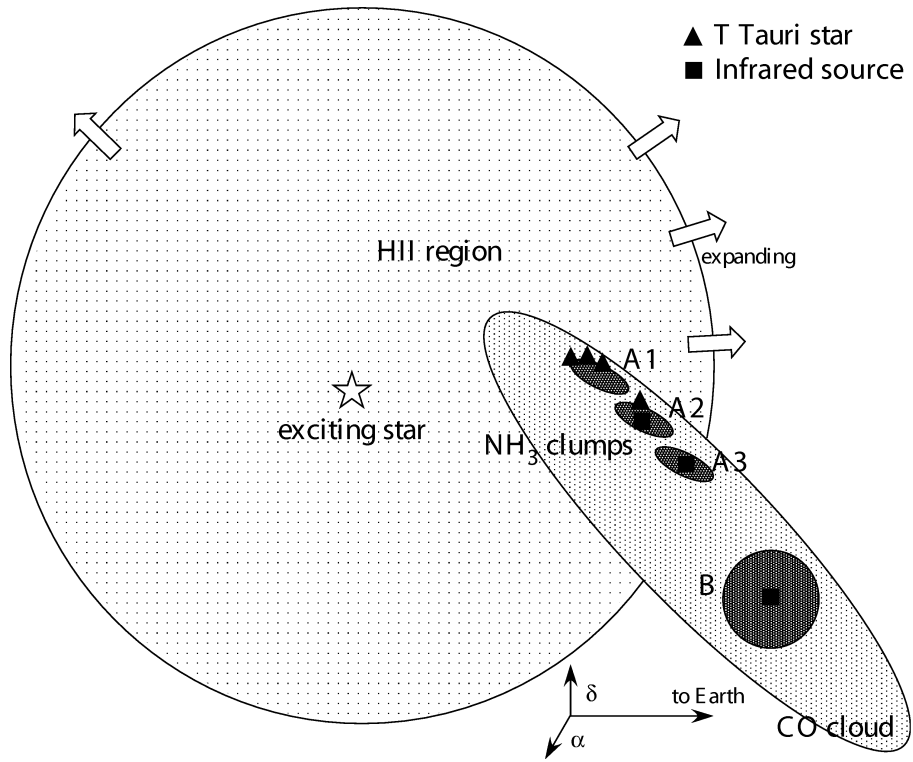}
\end{center}
\caption{Schematic geometry of the NH$_3$ clumps and the H\emissiontype{II} region.}
\label{fig:13}
\end{figure*}

In clump-B, the H\emissiontype{$\alpha$} emission is not detected.
This indicates that there is no ionized gas at the foreground of clump-B.
However, we found the presence of ionized gas in the background of clump-B,
because a radio continuum emission at 4.8 GHz is detected there \citep{wendker1984}.
Therefore, clump-B would be located in the foreground of the H\emissiontype{II} region.
We show the schematic geometry of the NH$_3$ clumps and the H\emissiontype{II} region
in Figure \ref{fig:13}.


\subsection{Why is the SFE of Subclump-A1 high?}

In subsection \ref{sec:4.2},
we show that the SFE $\simeq$ 36--62\% of subclump-A1 is 
higher than that of other molecular clouds.
Here, we examine why this is so.

There is no difference in 
the physical condition of the molecular gas in clump-A and clump-B.
The kinetic temperature and the velocity width of clump-A are similar
to those of clump-B.
The main difference between them appears 
to be the geometry to the H\emissiontype{II} region.
The geometry shown in the previous subsection suggests that 
subclump-A1 is 
closer to the H\emissiontype{II} region than any of the other subclumps or clump-B.
The five T-Tauri stars in subclump-A1 would be formed by 
the interaction of the H\emissiontype{II} region with the molecular gas.
A high SFE is observed in other triggered star forming regions.
The NGC 2024 and 2068 molecular clouds, each of which interact 
with an H\emissiontype{II} region (\cite{chandler1996}),
show SFE $\simeq 42$\% (\cite{lada1992}).
This means that the SFE of subclump-A1 increases because of 
the effect of the H\emissiontype{II} region.

Theoretical calculations suggest
that the SFE is in the range of 30--50\% of 
the regions that form clusters of low-mass stars \citep{matzner2000}.
The estimated SFE of subclump-A1 is very close to this value.
\citet{matzner2000} suggest that
the SFE of the clumps that are more massive than approximately 3000 \MO, 
in which O stars will form,
is lower than 30--50\% because of the destructive effects of massive stars.
The SFE may be increased in a cloud with the formation of a low-mass star cluster.

We consider whether other subclumps and clump-B are kept the low SFE.
Star formation appears to advance sequentially in the order of A1, A2, A3, and B.
This order is the same as that of the distance from the H\emissiontype{II} region.
It is suggested that
the triggered star formation or the interstellar shock comes
sequentially from the H\emissiontype{II} region.
A molecular shell with an expansion velocity of $\simeq 5$ km s$^{-1}$
is found in the $^{12}$CO line \citep{bally1980}.
In the case that the effect of the H\emissiontype{II} region expands at this velocity,
the crossing timescale from subclump-A1 to clump-B
is estimated to be $\simeq 6 \times 10^5$ yr using their separation of 2.9 pc.
This timescale is shorter than the lifetime of the O5 V type star
(2 $\times$ 10$^{6}$ yr; \cite{walborn2007}),
which is considered to be the ionizing star of the H\emissiontype{II} region
\citep{comeron2005}.
This suggests that the H\emissiontype{II} region can affect clump-B in the future,
in the case that the separation in the line of sight is
the same order of magnitude as that on the sky.
The total molecular gas mass of clump-A and clump-B is 391 \MO.
This mass is close to that of NGC 2024 or NGC 2068 \citep{lada1992},
and it is small enough to avoid cloud destruction by new born stars \citep{matzner2000}.
This suggests that the SFE of the combined clump-A and B 
can be as high as approximately 40\%.

The observed SFE is sensitive to the estimation of both gas and stellar masses.
As mentioned in subsection \ref{sec:3.2}, 
we estimated the actual gas mass within a factor of 2--3.
However, it is generally difficult to estimate the molecular gas mass.
The estimated molecular gas mass is sometimes different by more than a factor of 10
in different observed lines.
In the W3 giant molecular cloud, 
the molecular gas mass estimated in the $^{12}$CO ($J$=1-0), 
C$^{18}$O ($J$=2-1), and NH$_3$ lines is 
16000, 1400, and 3300 \MO, respectively \citep{tieftrunk1998},
although they were estimated in the same area.
For the estimation of SFE, 
the $^{12}$CO line data are often used to estimate the molecular gas mass 
(e.g. \cite{myers1986}; \cite{leisawitz1989}).
Because the $^{12}$CO line traces the less dense gas,
the molecular gas mass might be overestimated.
However, the data of the molecular line to trace the dense gas 
is not ideal to estimate the molecular gas mass,
because the relative abundance is difficult to determine precisely.
For example, the abundance of NH$_3$ varies by up to a factor of 10 from cloud to cloud.

It is also difficult to estimate the stellar mass accurately.
There are few studies based on total stellar mass 
estimated as the sum of the masses of individual stars.
Although the infrared luminosity is often used to estimate the stellar mass,
it would be less accurate than the mass estimated based on the number 
count of the T-Tauri type stars shown in this paper.

As mentioned above, the observed SFE reported in some studies 
should be revised by an order of magnitude.
The SFEs of the nearby molecular clouds such as
Perseus and Ophiuchus, L1551 in Taurus, and giant molecular clouds in the inner Galaxy
are 3--6\%, 9--15\%, and 2\%, respectively 
(\cite{myers1986}; \cite{swift2008}; \cite{evans2009}).
These values may be underestimated, 
because these studies use the molecular gas mass estimated from the $A_V$ and CO maps.
We should take care how to estimate the SFE to refer it from the previous studies.


\subsection{Future H$_2$O Maser Surveys}

\begin{figure*}[h]
\begin{center}
\FigureFile(80mm,100mm){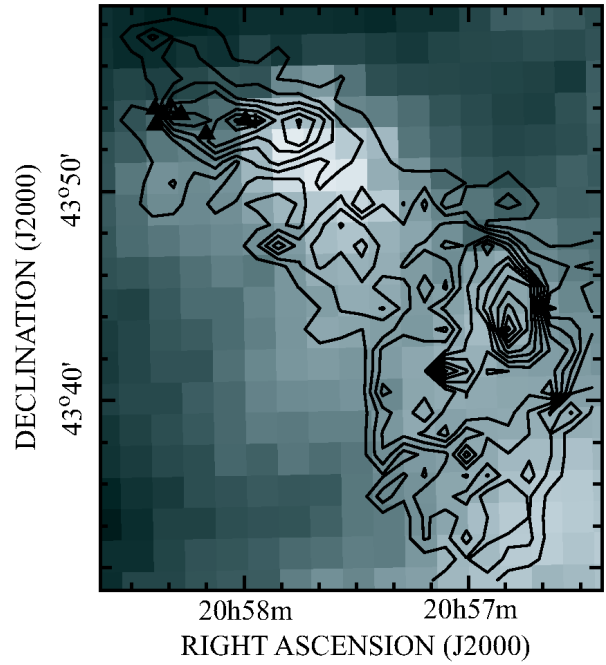}
\end{center}
\caption{IRAS 100 $\mu$m image (gray scale) on which the integrated intensity map of 
          NH$_3$ (1,1) line (contour) is superimposed.}
\label{fig:14}
\end{figure*}

Previous surveys of H$_2$O masers have been carried out based on the 
IRAS Point Source Catalogue (PSC).
This catalogue is useful for searching for YSOs embedded in the molecular clouds.
However, the counterpart of the H$_2$O maser 
that we found in subclump-A2 is not catalogued.
It shows that the H$_2$O survey based on the IRAS PSC is insufficient.
There are two possibilities why some YSOs are 
uncatalogued in the IRAS PSC:
sensitivity too poor to detect them or resolution too poor
to resolve a cluster of some sources.
Figure \ref{fig:14} shows an IRAS image at 100 $\mu$m 
overlayed on our NH$_3$ map.
A complex source is found near clump-A in the IRAS image, although it is composed of
several infrared sources in the $Spitzer$ image 
(see Figure \ref{fig:10}).

This suggests that there are many H$_2$O maser sources 
which are not catalogued in the IRAS PSC.
A new H$_2$O maser survey should be carried out based on a point source catalogue with 
a higher resolution and sensitivity, such asq $Spitzer$ and/or AKARI should be carried out. 
Our new H$_2$O maser is associated with a far-infrared source, and its
luminosity is brighter than that of the mid-infrared.
This characteristic may be a good criterion with which to find new H$_2$O maser sources.


\section{Conclusions}

We observed NGC 7000 in the NH$_3$ line and H$_2$O maser
using the Kashima 34-m telescope.
Our observations are summarized as follows:

\begin{enumerate}

\item
We found two major clumps with a mass of 95--452 \MO, 
and three subclumps with a mass of 9--20 \MO.
The molecular gas in these show 
similar gas kinetic temperatures
of 11--15 K and line width of 1--2 km s$^{-1}$.
However, they show different star formation activities
such as the concentration of T-Tauri type stars and 
the association of an H$_2$O maser.

\item
One of the clumps that is associated with a 
cluster of T-Tauri type stars shows the SFE $\simeq$ 36--62\%.
This SFE is higher than that of the other clumps.

\item
A comparison of the distribution of molecular gas
and ionized gas traced by the H$\alpha$ emission
suggests that
the clump with high SFE is located near the H\emissiontype{II} region.
Therefore, the high SFE would be related to 
the interaction of molecular gas and the H\emissiontype{II} region.

\item
We found a new H$_2$O maser source in the NH$_3$ clump.
Although the counterpart of this maser is not found in the IRAS point source catalogue,
we found it in the $Spitzer$ 24- and 70-$\mu$m images.
This suggests that
a new H$_2$O maser survey  should be carried out based on the point source catalogue of 
$Spitzer$ and/or AKARI. 

\end{enumerate}


\bigskip

We thank an anonymous referee for very useful comments and suggestions.
T.O. was supported by a Grant-in-Aid for Scientific Research
from the Japan Society for the Promotion Science (17340055).
We acknowledge K. Miyazawa (NAOJ) for his technical support of observations.



\clearpage

\begin{table*}[h]
\begin{center}
\caption{Line parameters obtained at the peak position of the clumps and the subclumps}
\label{tab:1}
\begin{tabular}{ccccrrrrr}
\hline \hline
\multicolumn{1}{c}{Clump}                                          &
\multicolumn{1}{c}{R.A.}                                           &
\multicolumn{1}{c}{Decl.}                                          &
\multicolumn{1}{c}{Line}                                           &
\multicolumn{1}{c}{$T_{\rm{MB}}$\footnotemark[$*$]}                &
\multicolumn{1}{c}{$v_{\rm{LSR}}$\footnotemark[$\dagger$]}         &
\multicolumn{1}{c}{$\Delta v$\footnotemark[$\dagger$]}             &
\multicolumn{1}{c}{$\int T_{\rm{MB}} dv$\footnotemark[$\ddagger$]} &
\multicolumn{1}{c}{rms noise}                                      \\
                                          &
\multicolumn{1}{c}{(J2000)}               &
\multicolumn{1}{c}{(J2000)}               &
\multicolumn{1}{c}{($J,K$)}               &
\multicolumn{1}{c}{(K)}                   &
\multicolumn{1}{c}{(km s$^{-1}$)}         &
\multicolumn{1}{c}{(km s$^{-1}$)}         &
\multicolumn{1}{c}{(K km s$^{-1}$)}       &
\multicolumn{1}{c}{(K)}                   \\
\hline
A1 & \timeform{20h58m18.8s} & \timeform{+43D53'24"} & (1,1) &       1.92                       &      5.6 &      1.5 & 2.96$\pm$0.13 & 0.10 \\
   &                         &                        & (2,2) &       0.43                       &      5.0 &      1.2 & 0.55$\pm$0.15 & 0.11 \\
   &                         &                        & (3,3) & $\leq$0.30\footnotemark[$\S$] & $\cdots$ & $\cdots$ & $\cdots$      & 0.10 \\
\\
A2 & \timeform{20h58m02.1s} & \timeform{+43D53'24"} & (1,1) &       2.59                      &      5.5 &      1.4 & 3.85$\pm$0.09 & 0.07 \\
   &                         &                        & (2,2) &       0.56                       &      5.4 &      1.6 & 0.93$\pm$0.11 & 0.08 \\
   &                         &                        & (3,3) & $\leq$0.21\footnotemark[$\S$] & $\cdots$ & $\cdots$ & $\cdots$      & 0.07 \\
\\
A3 & \timeform{20h57m45.5s} & \timeform{+43D53'24"} & (1,1) &       1.44                      &      5.3 &      2.1 & 3.21$\pm$0.13 & 0.08 \\
   &                         &                        & (2,2) &       0.38                       &      5.0 &      2.7 & 1.11$\pm$0.12 & 0.08 \\
   &                         &                        & (3,3) & $\leq$0.24\footnotemark[$\S$] & $\cdots$ & $\cdots$ & $\cdots$      & 0.08 \\
\\
B  & \timeform{20h56m49.9s} & \timeform{+43D43'24"} & (1,1) &       2.79                      &      1.5 &      1.9 & 6.25$\pm$0.21 & 0.14 \\
   &                         &                        & (2,2) &       0.70                       &      1.5 &      1.8 & 1.58$\pm$0.20 & 0.14 \\
   &                         &                        & (3,3) & $\leq$0.42\footnotemark[$\S$] & $\cdots$ & $\cdots$ & $\cdots$      & 0.14 \\
\hline
\multicolumn{9}{@{}l@{}} {\hbox to 0pt{\parbox{170mm}{\footnotesize
\footnotemark[$*$]
   The error of the Gaussian fitting is close to the rms noise level.
\par\noindent
\footnotemark[$\dagger$]
   The error of the Gaussian fitting is much smaller than the velocity resolution (0.39 km s$^{-1}$).
\par\noindent
\footnotemark[$\ddagger$]
   The error corresponds to one standard deviation.
\par\noindent
\footnotemark[$\S$]
   The upper limit is given as 3 times of the rms noise.
\par\noindent
}\hss}}
\end{tabular}
\end{center}
\end{table*}

\begin{table*}[h]
\begin{center}
\caption{Physical properties of the clumps and the subclumps}
\label{tab:2}
\begin{tabular}{crrrrrr}
\hline \hline
\multicolumn{1}{c}{Clump}          &
\multicolumn{1}{c}{Size}           &
\multicolumn{1}{c}{$\tau_{(1,1)}$} &
\multicolumn{1}{c}{$T_{\rm{rot}}$} &
\multicolumn{1}{c}{$N$(NH$_3$)}    &
\multicolumn{1}{c}{$M_{\rm{LTE}}$} &
\multicolumn{1}{c}{$M_{\rm{vir}}$} \\
                                   &
\multicolumn{1}{c}{(pc)}           &
                                   &
\multicolumn{1}{c}{(K)}            &
\multicolumn{1}{c}{(cm$^{-2}$)}    &
\multicolumn{1}{c}{(\MO)}          &
\multicolumn{1}{c}{(\MO)}          \\
\hline
A1 & 0.21 & 1.4$\pm$0.4 & 12$\pm$2 & 2.4$\times$10$^{15}$ &  12 & $\cdots$ \\
A2 & 0.31 & 1.2$\pm$0.3 & 13$\pm$1 & 1.8$\times$10$^{15}$ &  20 & $\cdots$ \\
A3 & 0.21 & 0.8$\pm$0.4 & 15$\pm$2 & 1.8$\times$10$^{15}$ &   9 & $\cdots$ \\
A  & 0.67 & 1.2$\pm$0.2 & 13$\pm$1 & 1.8$\times$10$^{15}$ &  95 & 125      \\
B  & 1.02 & 1.6$\pm$0.2 & 11$\pm$1 & 3.7$\times$10$^{15}$ & 452 & 350      \\
\hline
\end{tabular}
\end{center}
\end{table*}

\begin{table*}[h]
\begin{center}
\caption{Star formation efficiency of individual clumps}
\label{tab:3}
\begin{tabular}{crrrrrr}
\hline \hline
\multicolumn{1}{c}{Clump}              &
\multicolumn{2}{c}{Number of sources}  &
\multicolumn{2}{c}{Stellar mass (\MO)} &
\multicolumn{2}{c}{SFE (\%)}           \\
\multicolumn{1}{c}{}                   &
\multicolumn{2}{c}{\hrulefill \ }     &
\multicolumn{2}{c}{\hrulefill \ }     &
\multicolumn{2}{c}{\hrulefill \ }     \\
\multicolumn{1}{c}{}                   &
\multicolumn{1}{c}{T-Tauri}            &
\multicolumn{1}{c}{Total}              &
\multicolumn{1}{c}{T-Tauri}            &
\multicolumn{1}{c}{Total}              &
\multicolumn{1}{c}{T-Tauri}            &
\multicolumn{1}{c}{Total}              \\
\hline
A1 & 5 & $<15$ &   6.9 & $<19.9$ & 36 & $<62$ \\
A2 & 1 & $ <5$ &   0.8 & $ <6.0$ &  4 & $<23$ \\
A3 & 0 & $ <4$ &     0 & $ <5.2$ &  0 & $<36$ \\
A  & 7 & $<32$ &   7.7 & $<41.5$ &  8 & $<30$ \\
B  & 0 & $<24$ &     0 & $<31.2$ &  0 & $< 6$ \\
\hline
\end{tabular}
\end{center}
\end{table*}


\clearpage

\begin{thebibliography}{}
\bibitem[Anglada et al.(1994)]{anglada1994} 
Anglada, G., Rodriguez, 
L.~F., Girart, J.~M., Estalella, R., 
\& Torrelles, J.~M.\ 1994, \apjl, 420, L91 
\bibitem[Aspin(2003)]{aspin2003} Aspin, C.\ 2003, \aj, 125, 1480 
\bibitem[Bachiller(1996)]{bachiller1996} Bachiller, R.\ 1996, \araa, 34, 111 
\bibitem[Bally \& Scoville(1980)]{bally1980} 
Bally, J., \& Scoville, N.~Z.\ 1980, \apj, 239, 121 
\bibitem[Bohlin et al.(1978)]{bohlin1978} 
Bohlin, R.~C., Savage, 
B.~D., \& Drake, J.~F.\ 1978, \apj, 224, 132 
\bibitem[Cambr{\'e}sy et al.(2002)]{cambresy2002} 
Cambr{\'e}sy, L., 
Beichman, C.~A., Jarrett, T.~H., \& Cutri, R.~M.\ 2002, \aj, 123, 2559 
\bibitem[Chandler \& Carlstrom(1996)]{chandler1996} 
Chandler, C.~J., \& Carlstrom, J.~E.\ 1996, \apj, 466, 338 
\bibitem[Claussen et al.(1996)]{claussen1996} 
Claussen, M.~J., Wilking, B.~A., Benson, P.~J., Wootten, A., Myers, P.~C., 
\& Terebey, S.\ 1996, \apjs, 106, 111 
\bibitem[Codella et al.(1997)]{codella1997} 
Codella, C., Welser, R., Henkel, C., Benson, P.~J., \& Myers, P.~C.\ 1997, \aap, 324, 203 
\bibitem[Cohen \& Kuhi(1979)]{cohen1979} 
Cohen, M., \& Kuhi, L.~V.\ 1979, \apjs, 41, 743 
\bibitem[Comer{\'o}n \& Pasquali(2005)]{comeron2005} 
Comer{\'o}n, F., \& Pasquali, A.\ 2005, \aap, 430, 541 
\bibitem[Dahm \& Simon(2005)]{dahm2005} 
Dahm, S.~E., \& Simon, T.\ 2005, \aj, 129, 829 
\bibitem[Danby et al.(1988)]{danby1988} 
Danby, G., Flower, D.~R., 
Valiron, P., Schilke, P., \& Walmsley, C.~M.\ 1988, \mnras, 235, 229 
\bibitem[Deharveng et al.(2003)]{deharveng2003} 
Deharveng, L., Lefloch, B., Zavagno, A., Caplan, J., Whitworth, A.~P., 
Nadeau, D., \& Mart{\'{\i}}n, S.\ 2003, \aap, 408, L25 
\bibitem[Deharveng et al.(2005)]{deharveng2005} 
Deharveng, L., Zavagno, A., \& Caplan, J.\ 2005, \aap, 433, 565 
\bibitem[Dobashi et al.(1994)]{dobashi1994} 
Dobashi, K., Bernard, 
J.-P., Yonekura, Y., \& Fukui, Y.\ 1994, \apjs, 95, 419 
\bibitem[Dobashi et al.(2001)]{dobashi2001} 
Dobashi, K., Yonekura, 
Y., Matsumoto, T., Momose, M., Sato, F., Bernard, J.-P., 
\& Ogawa, H.\ 2001, \pasj, 53, 85 
\bibitem[Duerr et al.(1982)]{duerr1982} 
Duerr, R., Imhoff, C.~L., 
\& Lada, C.~J.\ 1982, \apj, 261, 135 
\bibitem[Evans et al.(2009)]{evans2009} 
Evans, N.~J., et al.\ 
2009, \apjs, 181, 321 
\bibitem[Fountain et al.(1983)]{fountain1983} 
Fountain, W.~F., Gary, 
G.~A., \& Odell, C.~R.\ 1983, \apj, 269, 164 
\bibitem[Furuya et al.(2003)]{furuya2003} 
Furuya, R.~S., Kitamura, 
Y., Wootten, A., Claussen, M.~J., \& Kawabe, R.\ 2003, \apjs, 144, 71 
\bibitem[Genzel et al.(1982)]{genzel1982} Genzel, R., Ho, 
P.~T.~P., Bieging, J., \& Downes, D.\ 1982, \apjl, 259, L103 
\bibitem[Herbig(1958)]{herbig1958} 
Herbig, G.~H.\ 1958, \apj, 128, 259 
\bibitem[Ho \& Townes(1983)]{ho1983} 
Ho, P.~T.~P., \& Townes, C.~H.\ 1983, \araa, 21, 239 
\bibitem[Gandolfi et al.(2008)]{gandolfi2008} 
Gandolfi, D., et al.\ 
2008, \apj, 687, 1303 
\bibitem[Kun et al.(2009)]{kun2009} Kun, M., Balog, Z., Kenyon, 
S.~J., Mamajek, E.~E., \& Gutermuth, R.~A.\ 2009, \apjs, 185, 451 
\bibitem[Kutner \& Ulich(1981)]{kutner1981} 
Kutner, M.~L., \& Ulich, B.~L.\ 1981, \apj, 250, 341
\bibitem[Lada(1992)]{lada1992} 
Lada, E.~A.\ 1992, \apjl, 393, L25  
\bibitem[Lada et al.(1997)]{lada1997} 
Lada, E.~A., Evans, N.~J., 
II, \& Falgarone, E.\ 1997, \apj, 488, 286 
\bibitem[Ladd et al.(1994)]{ladd1994} 
Ladd, E.~F., Myers, P.~C., 
\& Goodman, A.~A.\ 1994, \apj, 433, 117 
\bibitem[Laugalys \& Strai{\v z}ys(2002)]{laugalys2002} 
Laugalys, V., \& Strai{\v z}ys, V.\ 2002, Baltic Astronomy, 11, 205 
\bibitem[Lang \& Willson(1980)]{lang1980} 
Lang, K.~R., \& Willson, R.~F.\ 1980, \apj, 238, 867 
\bibitem[Leisawitz et al.(1989)]{leisawitz1989} 
Leisawitz, D., Bash, 
F.~N., \& Thaddeus, P.\ 1989, \apjs, 70, 731 
\bibitem[MacLaren et al.(1988)]{maclaren1988} MacLaren, I., 
Richardson, K.~M., \& Wolfendale, A.~W.\ 1988, \apj, 333, 821 
\bibitem[Matzner \& McKee(2000)]{matzner2000} 
Matzner, C.~D., \& McKee, C.~F.\ 2000, \apj, 545, 364 
\bibitem[Myers \& Benson(1983)]{myers1983} 
Myers, P.~C., \& Benson, P.~J.\ 1983, \apj, 266, 309 
\bibitem[Myers et al.(1986)]{myers1986} 
Myers, P.~C., Dame, 
T.~M., Thaddeus, P., Cohen, R.~S., Silverberg, R.~F., Dwek, E., 
\& Hauser, M.~G.\ 1986, \apj, 301, 398 
\bibitem[Nakajima et al.(2007)]{nakajima2007} 
Nakajima, T., et al.\ 
2007, \pasj, 59, 1005
\bibitem[Prasad \& Huntress(1980)]{prasad1980} 
Prasad, S.~S., \& Huntress, W.~T., Jr.\ 1980, \apjs, 43, 1 
\bibitem[Ridge et al.(2003)]{ridge2003} 
Ridge, N.~A., Wilson, T.~L., Megeath, S.~T., Allen, L.~E., \& Myers, P.~C.\ 2003, \aj, 126, 286 
\bibitem[Rieke \& Lebofsky(1985)]{rieke1985} 
Rieke, G.~H., \& Lebofsky, M.~J.\ 1985, \apj, 288, 618 
\bibitem[Rohlfs \& Wilson(1996)]{rohlfs1996} 
Rohlfs, K., \& Wilson, T.~L.\ 1996, Tools of Radio Astronomy, XVI, 
423 pp.~127 figs., 20 tabs..~ Springer-Verlag Berlin Heidelberg New York.~Also Astronomy and Astrophysics Library,  
\bibitem[Strai{\v z}ys \& Laugalys(2008)]{straizys2008} 
Strai{\v z}ys, V., \& Laugalys, V.\ 2008, Baltic Astronomy, 17, 143 
\bibitem[Sugitani et al.(1989)]{sugitani1989} 
Sugitani, K., Fukui, 
Y., Mizuni, A., \& Ohashi, N.\ 1989, \apjl, 342, L87 
\bibitem[Sugitani et al.(1991)]{sugitani1991} 
Sugitani, K., Fukui, 
Y., \& Ogura, K.\ 1991, \apjs, 77, 59 
\bibitem[Sugitani \& Ogura(1994)]{sugitani1994} 
Sugitani, K., \& Ogura, K.\ 1994, \apjs, 92, 163 
\bibitem[Swift \& Welch(2008)]{swift2008} 
Swift, J.~J., \& Welch, W.~J.\ 2008, \apjs, 174, 202 
\bibitem[Tieftrunk et al.(1998)]{tieftrunk1998} 
Tieftrunk, A.~R., Megeath, S.~T., Wilson, T.~L., \& Rayner, J.~T.\ 1998, \aap, 336, 991 
\bibitem[Ungerechts et al.(1980)]{ungerects1980} 
Ungerechts, H., Walmsley, C.~M., \& Winnewisser, G.\ 1980, \aap, 88, 259 
\bibitem[Ungerechts et al.(1982)]{ungerects1982} 
Ungerechts, H., Winnewisser, G., \& Walmsley, C.~M.\ 1982, \aap, 111, 339 
\bibitem[Walborn(2007)]{walborn2007} 
Walborn, N.~R.\ 2007, 
arXiv:astro-ph/0701573 
\bibitem[Walmsley \& Ungerechts(1983)]{walmsley1983} 
Walmsley, C.~M., \& Ungerechts, H.\ 1983, \aap, 122, 164 
\bibitem[Wendker(1984)]{wendker1984} 
Wendker, H.~J.\ 1984, \aaps, 58, 291 
\end{thebibliography}
\end{document}